\documentclass{jfm}
\usepackage{graphicx}
\usepackage{epstopdf, epsfig}
\usepackage{amsmath}
\usepackage{subfig}
\usepackage{xcolor}
\usepackage{soul}

\def\Rey {{Re}}
\def\Retau {{Re_{\tau}}}
\def\reffig#1{Fig.~\ref{fig:#1}}

\def\refapp#1{Appendix~\ref{app:#1}}
\def\refeq#1{Eq.~(\ref{eq:#1})}

\def\vel {\mathbf{u}}
\def\basevel {\mathbf{U}}

\shorttitle{Near-wall invariant solution}
\shortauthor{S. Azimi and T. M. Schneider}

\title{Self-similar invariant solution in the near-wall region of a turbulent boundary layer  at asymptotically high Reynolds numbers}

\author{Sajjad Azimi\aff{1}
\and Tobias M.  Schneider\aff{1}
\corresp{\email{tobias.schneider@epfl.ch}}}

\affiliation{\aff{1}Emergent Complexity in Physical Systems Laboratory (ECPS), \'Ecole Polytechnique F\'ed\'erale de Lausanne, CH-1015 Lausanne, Switzerland}

\begin{document}

\maketitle

\begin{abstract}
At sufficiently high Reynolds numbers, shear-flow turbulence close to a wall acquires universal properties. When length and velocity are rescaled by appropriate characteristic scales of the turbulent flow and thereby measured in \emph{inner units}, the statistical properties of the flow become independent of the Reynolds number.
We demonstrate the existence of a wall-attached non-chaotic exact invariant solution of the fully nonlinear 3D Navier-Stokes equations for a parallel boundary layer that captures the characteristic self-similar scaling of near-wall turbulent structures. The branch of travelling wave solutions can be followed up to $\Rey=1,000,000$. Combined theoretical and numerical evidence suggests that the solution is asymptotically self-similar and exactly scales in inner units for Reynolds numbers tending to infinity. 
Demonstrating the existence of invariant solutions that capture the self-similar scaling properties of turbulence in the near-wall region is a step towards extending the dynamical systems approach to turbulence from the transitional regime to fully developed boundary layers.

\end{abstract}


%%%%%%%%%%%%%%%%%%%%%%%%%%%%%%%%%%%%%%%%%%%%%%%%%%%%%%%%%%%%%%%%%%%%%%%%%%%%%%%%%%%%%%%%%%%%%%%
\section{Introduction}\label{sec:intro}
% MOTIVATION - maybe reword a bit.. very similar to Gibson & Schneider 2016
Invariant solutions of the fully nonlinear 3D Navier-Stokes equations are known to play an important role in the dynamics of turbulence at low Reynolds numbers. For virtually all canonical shear flows, invariant solutions in the form of equilibria, travelling waves and periodic orbits, have been computed. The solutions act as transiently visited building blocks for the dynamics \citep{Gibson2007, Kawahara2012, Suri2017} and capture many characteristic features of transitional flows including self-organised turbulent-laminar patterns such as puffs in pipe flow and laminar-turbulent stripes in Couette flow \citep{Avila2013, Reetz2019a}. 

To extend the approach to describe turbulence in terms of invariant-solutions from the transitional regime to developed turbulent wall-bounded flows at high Reynolds numbers relevant for many engineering applications, invariant solutions capturing the characteristics of those fully turbulent boundary layer flows are required.
At sufficiently high flow speeds, turbulent fluctuations in a layer close to the wall show a typical spacing of streaky motions that is universal and independent of the specific flow parameters, when distances are measured in inner or wall units \citep{Kline1967, Kim1987}. The wall shear stress  $\tau_w$ controls the characteristic scales of turbulence, namely the friction velocity $u_\tau=\sqrt{\tau_w / \rho}$ and the viscous length unit $\delta_\tau = \nu / u_\tau$. After rescaling velocities and distances with these characteristic scales, the turbulent flow in the \emph{inner region} close to the wall becomes independent of the Reynolds number \citep{Jimenez2018}.

The self-similar inner region interacts with the \emph{outer region} further away from the wall. Here turbulent fluctuations do not scale in inner units but change with Reynolds number. The characteristic length scale of the outer region $l_{out}$ depends on the specific flow system. For a semi-infinite open flow domain the outer scale is given by the boundary layer thickness of the turbulent flow. In confined flows, such as channel flow, the turbulent boundary layer cannot freely expand so that confinement effects limit the outer scale to the distance between the walls. The higher the flow speed and thus wall shear, the thinner is the inner region.
The friction Reynolds number $Re_\tau = u_\tau l_{out} / \nu = l_{out}/\delta_\tau$ measures the scale separation between the self-similar near-wall inner scale and the characteristic outer scale of the turbulent flow. $\Retau$ thereby characterises fully developed turbulence along a wall and indicates the strength of turbulence.

To capture the universal small-scale motions of turbulence in the inner region close to the wall, invariant solutions are required that are localised at the wall, exist at very high Reynolds numbers and scale in inner units defined by the mean wall-shear stress of turbulence. However, to date, attempts to find invariant solutions of the Navier-Stokes equations capturing the universal features of the small-scale motions in the near-wall region have mostly failed. \cite{Rawat2015a} were unable to find a near-wall solution connected to the Nagata equilibrium \citep{Nagata1990}, the solution of \cite{Jimenez2001} requires non-physical artificial damping and the wall-attached solution of \cite{Neelavara2017} fails to scale in inner units.
\cite{Deguchi2015b} identifies a solution which scales in inner units at high Reynolds numbers but is not localised at the wall. More recently \cite{Eckhardt2017} present two solutions in plane Couette flow, one localized in the center of the channel and one attached to the wall. The solutions are followed up to a Couette Reynolds number of $Re=100,000$, and become approximately Reynolds number independent when rescaled by the inner length scale. Likewise \cite{Yang2019} follow a wall-attached solution in fixed flux channel flow up to $Re_\tau=268$ and show that the solution approximately scales in inner units. Both \cite{Eckhardt2017} and \cite{Yang2019} use inner units corresponding to the wall-drag of the solution itself, which differs from the mean wall drag of turbulence at the same controlled relative plate velocity in Couette or controlled flux in channel flow.

%[WE DO!]
Here we present a wall-attached solution of a parallel boundary layer at Reynolds numbers up to $1,000,000$. For large Reynolds numbers, the solution scales in inner units based on the mean turbulent wall drag. Combined numerical and theoretical evidence suggest an exactly self-similar solution, that is asymptotically independent of Reynolds number when rescaled in terms of inner units. The wall-attached solution thus captures the characteristic scaling behaviour of the near-wall turbulence universally observed in wall-bounded flows at high flow speeds.  

%%%%%%%%%%%%%%%%%%%%%%%%%%%%%%%%%%%%%%%%%%%%%%%%%%%%%%%%%%%%%%%%%%%%%%%%%%%%%%%%%%%%%%%%%%%%%%%
\section{The asymptotic suction boundary layer}\label{sec:ASBL}
\begin{figure}
    \centering
    {\includegraphics[width=0.4\textwidth]{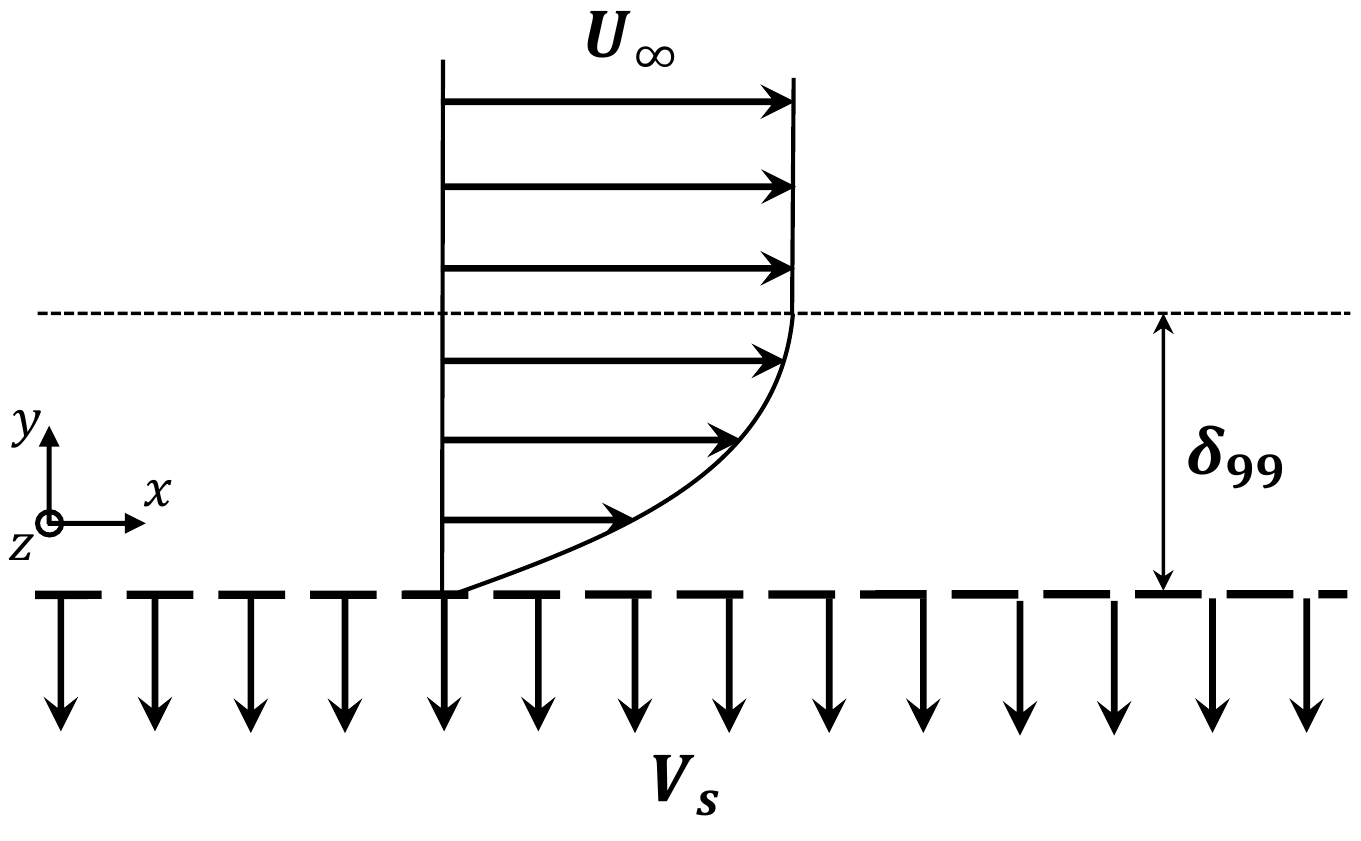}}
    \caption{Schematic of the asymptotic suction boundary layer flow (ASBL). The streamwise, wall-normal and spanwise directions are denoted by $x$, $y$, and $z$, the corresponding velocities are denoted by $u$, $v$ and $w$, respectively. The height where the mean streamwise velocity is $99\%$ of the free stream velocity is called the boundary layer thickness and denoted by $\delta_{99}$.}
    \label{fig:schematic}
\end{figure}

Previous attempts to identify high Reynolds number invariant solutions in the near-wall region have considered confined shear flows such as channel flow. Owing to the universal nature of near-wall turbulence at sufficiently high Reynolds number we instead consider a boundary layer in a semi-infinite domain. To avoid the complications associated to non-parallelism of the flow we study the fully turbulent asymptotic suction boundary layer (ASBL), where moderate suction at the wall keeps the base flow parallel in the downstream asymptotic regime \citep{Schlichting2004}.

We consider the flow developing over a flat plate immersed in a uniform stream of velocity $U_\infty$ with constant and uniform suction $V_s$ into the plate (\reffig{schematic}). 
Sufficiently far downstream, where wall friction balances streamwise momentum loss due to the suction, the boundary layer thickness reaches a constant value and `asymptotic suction boundary layer' flow is reached.
The laminar exact solution of the Navier-Stokes equations is $\check{U}/U_\infty=1-\exp(-\check{y} V_s/\nu)$, where $\check{U}$ is the (dimensional) streamwise velocity, $\check{y}$ the (dimensional) wall-normal coordinate and $\nu$ is the kinematic viscosity of the fluid.
This laminar solution is characterised by a constant displacement thickness $\delta^*=\nu/V_s$ and can be recast in dimensionless form $U=1-e^{-y}$ by using $\delta^*$ as reference length and $U_\infty$ as reference velocity.
We prescribe the Reynolds number based on the laminar boundary layer displacement thickness $\Rey=U_\infty \delta^*/\nu = U_\infty/V_s $. This Reynolds number, referred as `the Reynolds number' is commonly used as control parameter for ASBL and needs to be distinguished from the friction Reynolds number $\Retau$. Despite its linear stability up to $\Rey=54370$ \citep{Hocking1975}, the laminar ASBL solution is in practice only observed for $\Rey \lesssim 270$ \citep{Khapko2016}; above this value the flow is turbulent.

There is numerous experimental, numerical and theoretical support for the fact that close to a wall, high Reynolds number turbulence is universal and independent of the specific system in which it is observed \citep{Pope2000, Jimenez2018}. At high $\Rey$ the universal near-wall turbulent dynamics in the small inner region decouples from large scale flow in the outer region. Since the universal features of near-wall turbulence can be studied in any wall-bounded shear flow, one may choose to consider a specific flow based on convenient properties of the non-universal outer scale dynamics.

ASBL flow has two key properties that are advantageous for studying near-wall turbulence when compared to other commonly studied canonical flows such as the fixed-flux channel flow considered by \cite{Yang2019}:

(1) Inner length and velocity scales capture properties of the turbulent state and are thus in general not known a priori. In ASBL, however, momentum conservation allows to directly control the wall drag so that the characteristic scale of near-wall turbulence can be expressed directly in terms of the control parameters of the flow. Consequently, one can carry out numerical studies in a domain whose size is fixed in inner units of the turbulent state. For some other canonical flows, this is not possible: For channel flow with constant flux, as studied by \cite{Yang2019}, the imposed control parameter is the Reynolds number based on the mean flow rate. A turbulent simulation or experiment is required to determine the wall shear of the turbulence $\tau_w$ and quantities can be rescaled with the inner velocity scale $u_\tau=\sqrt{\tau_w / \rho}$ and the inner length scale $\delta_\tau = \nu / u_\tau$ only a posteriori. If not the flux but the applied pressure gradient were fixed, also in channel flow, inner length and velocity scales could be determined a priori.

(2) For a confined shear flow such as channel flow where the outer scale is given by the separation of two bounding walls, numerically resolving a near-wall solution at high $Re_\tau$ in general implies to numerically resolve the entire flow domain including both walls with sufficient resolution to handle the large separation between outer and inner scales. For ASBL, localisation of the flow at a single wall allows to not fully resolve outer scales but focus on the near-wall region.

\section{Determining the minimal flow unit in inner units}

We consider a numerical domain of length $L_x$, width $L_z$ and height $H$, where periodic boundary conditions are applied in the $x$ and $z$ directions. On the upper and lower boundary, we impose Dirichlet conditions $\vel(x,0,z)=\vel(x,H,z)=\mathbf{0}$, with $\vel$ the deviation from the laminar solution. ASBL has continuous translational symmetries in $x$ and $z$ so that the periodic boundary conditions are compatible with the equivariance group of the flow problem. Consequently, any solution found in the periodic domain also exists in the infinitely extended system.
Based on turbulent simulations, we choose $L_x$ and $L_z$ such that small scale near-wall motions are faithfully captured. This defines the minimal flow unit (MFU). The height $H$ is chosen to be large enough so that the flow detaches from the top boundary and becomes independent of the domain height.

To determine the MFU in ASBL, we extract the most energetic length scales of the near-wall region from energy spectra.
A turbulent ASBL at $\Rey=333$ is simulated in a large domain of size $L_x=243, H=225, L_z=121.5$ (similar to simulations of \cite{Schlatter2011} and \cite{Bobke2016}; more details about the simulation are provided in \refapp{methods}).
The premultiplied streamwise energy spectrum peaks at $y^+=17$, where the plus superscript indicates quantities measured in terms of inner units. At this wall-normal location, the peaks in the streamwise and spanwise premultiplied energy spectra are located at $\lambda_x^+=633$ and $\lambda_z^+ = 170$, respectively. 
We consequently choose the length and the width of the MFU as $L_x^+=633$ and $L_z^+ = 170$ in inner units. This ensures that the most energetic modes of the near-wall region are captured.
A height of $H^+=632$ is sufficient to guarantee the complete detachment of all flow structures from the upper wall, as shown in \refapp{methods}. Note that the required height of the flow domain remains considerably smaller than the turbulent boundary layer thickness $\delta_{99}$ defining the outer scale. As discussed in \cite{Bobke2016}, this is related to the limited width of the MFU, disallowing the formation of the large-scale structures that extend far into the outer region of the turbulent boundary layer. The MFU capturing small scale near-wall motions thus has a size of  $L_x^+=633, H^+=632, L_z^+=170$ in inner units. Since the inner-unit location of the near-wall energy peak is independent of the Reynolds number, reflecting the inner-unit scaling of turbulence, the inner-unit box size determined here for $\Rey=333$ remains unchanged for higher Reynolds numbers.

Momentum conservation in the ASBL requires that the ratio of the friction velocity to the free stream velocity is  $u_\tau/U_\infty=1/\sqrt{\Rey}$, where $u_\tau = \sqrt{\tau_w/\rho}$, with $\tau_w$ the mean wall-friction. 
The viscous unit is $\delta_\tau = \delta^*/\sqrt{Re}$ where $\delta_\tau =\nu/u_\tau$ and $\delta^*$ is the displacement thickness of the laminar flow. Consequently, in ASBL both the inner velocity scale $u_\tau$ and the inner length scale $\delta_\tau$ are directly given in terms of the externally controlled Reynolds number and do not need to be computed from turbulent statistics. Thus we can exactly prescribe the size of a computational box in inner units, and continue invariant solutions towards high Reynolds numbers. As we increase $\Rey$, we change the size of the minimal flow unit in outer units in such a way that the size remains exactly constant in inner units.

%%%%%%%%%%%%%%%%%%%%%%%%%%%%%%%%%%%%%%%%%%%%%%%%%%%%%%%%%%%%%%%%%%%%%%%%%%%%%%%%%%%%%%%%%%%%%%%
\section{Invariant solutions in the minimal flow unit}\label{sec:MFU}
We aim at finding a wall-attached invariant solution that can be followed to very high $\Rey$. Instead of computing a `starting' solution at low Reynolds numbers close to the transition, where interaction with large-scale features of the flow, and in particular interaction between the two walls in Couette and Poiseuille flows, might have prevented the finding of a genuine `one-wall' solution in the near-wall inner region to start the continuation from, we immediately consider a value of $\Rey=1000$, well above transition to avoid any potential low Reynolds number effect in the selection of the solution branch. Moreover, in ASBL the kinetic energy associated with the non-universal and system-dependent large-scale features of the flow is relatively weak when compared to other flows \citep{Schlatter2011, Bobke2016}. Consequently, even at moderate Re, the flow dynamics is dominated by near-wall dynamics which may aid in identifying a solution branch representing a universal wall-attached high $\Rey$ solution.
At $\Rey=1000$, edge tracking \citep{Skufca2006, Schneider2007b} within the mirror symmetry subspace $[u,v,w](x,y,-z)=[u,v,-w](x,y,z)$ yields a travelling wave solution in the MFU determined above. 

%=================== CONTINUATION IN REYNOLDS NUMBER
The invariant solution computed at $\Rey=1000$ is used as starting point for a continuation in Reynolds number where the size of the domain is kept constant in inner units, and therefore shrinks in outer units, when $\Rey$ is increased.
As shown in \reffig{cont_in}, we can continue the solution up to $\Rey=1,000,000$ with both size of the box and magnitude of the solution decreasing in outer units. \reffig{CONTOUR} shows contours of the streamwise-averaged wall-normal velocity of the invariant solution in inner units at different Reynolds numbers. The solution structures remain almost unchanged over a wide range of Reynolds number when expressed in inner units.

To provide context for the achieved value of $\Rey$, we characterise the scale separation of fully developed turbulent flow at the same imposed value of $\Rey$ and in a large domain. While ASBL has the advantage that the self-similar near-wall inner scale $\delta_\tau$ is directly controlled by $\Rey$, determining the outer scale, commonly associated with the turbulent boundary layer thickness $\delta_{99}$, requires extensive turbulent simulations. To be domain-independent, these simulations need to be carried out in domains considerably larger than the MFU to allow for structures in the outer region including those large compared to the wavelength imposed by the periodic MFU box to develop. Those simulations can only be carried out for moderate $\Rey$. Using large eddy simulations, \cite{Schlatter2011, Bobke2016} determined the turbulent boundary layer thickness of ASBL as $\delta_{99} = 101$ for $Re=333$ and $\delta_{99} = 290$ for $Re=400$. \cite{Khapko2016} observed $\delta_{99}$ to grow linearly in Re for values between 260 and 333. For higher $\Rey$, the evolution of the turbulent boundary layer thickness is unknown. Since $\delta_\tau=0.001$ at the highest  achieved Reynolds number, the fully-developed turbulent flow at $\Rey=10^{6}$ has a scale separation of at least $290,000$, even if the growth of the turbulent boundary layer thickness does not extend much beyond $\Rey \sim 400$.

\begin{figure}
    \centering
    {\includegraphics[width=0.8\textwidth]{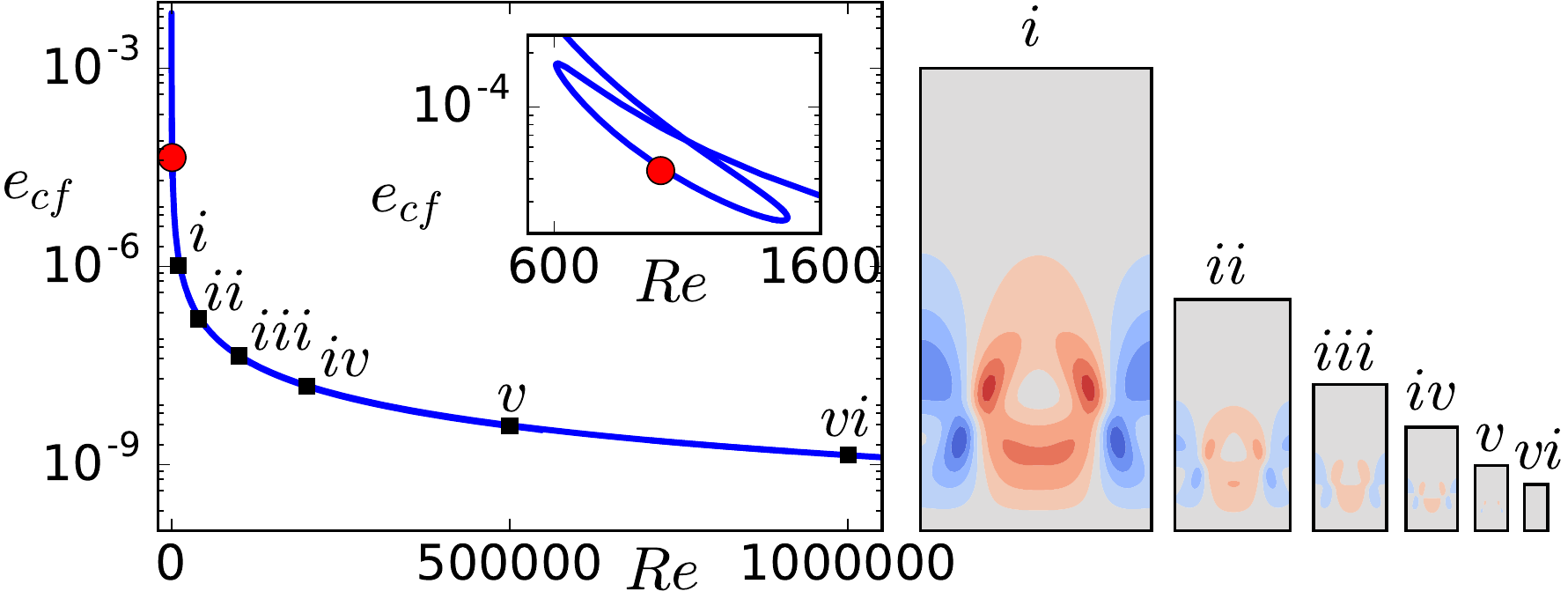}}
    \caption{Continuation of the invariant solution to high Reynolds numbers in outer units. Cross-flow kinetic energy per area of the wall $e_{cf}=\frac{1}{L_x L_z}\int_{MFU} \left(v^2 + w^2\right) dx dy dz$ in the MFU as a function of $\Rey$. Note that in outer units velocities are non-dimensionalised by $U_\infty$ and lengths by $\nu/V_s$, so $e_{cf}$ is in units of $\frac{U^2_\infty \nu}{V_s}$. The red dot indicates where the solution has been identified by edge tracking (see also inset for a magnification of the relevant parameter range). Visualisations of the invariant solution show contours of the streamwise-averaged wall-normal velocity at  $\Rey=10,000$, $40,000$, $100,000$, $200,000$, $500,000$, and $1,000,000$.}
    \label{fig:cont_in}
\end{figure}

\begin{figure}
    \centering
    \includegraphics[width=1\textwidth]{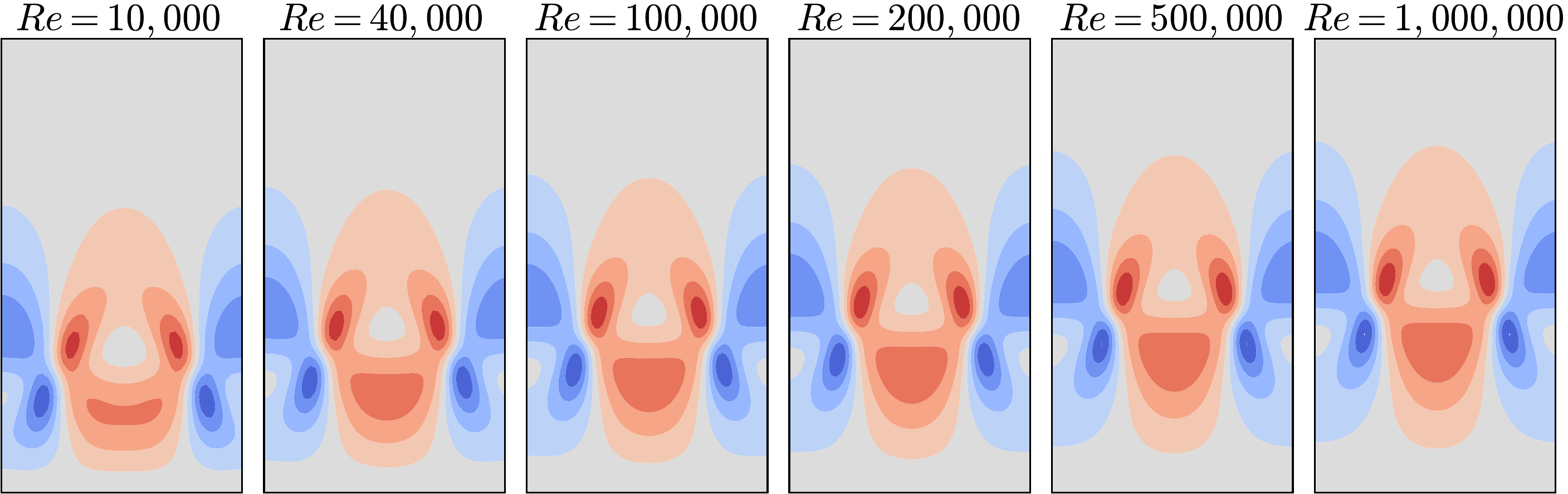}
    \caption{Invariant solution visualised in inner units. Contours of streamwise-averaged wall-normal velocity, normalised by friction velocity $u_\tau$ at the same values of $\Rey$  as in \reffig{cont_in}. The contour levels are $v^+ = v/u_\tau = \{\pm 0.015, \pm 0.045, \pm 0.075, \pm 0.105, \pm 0.135\}$. The solution is localised at the wall so that only the lower half of the numerical domain is shown.}
    \label{fig:CONTOUR}
\end{figure}

\begin{figure}
    \centering
	\includegraphics[width=0.3\textwidth]{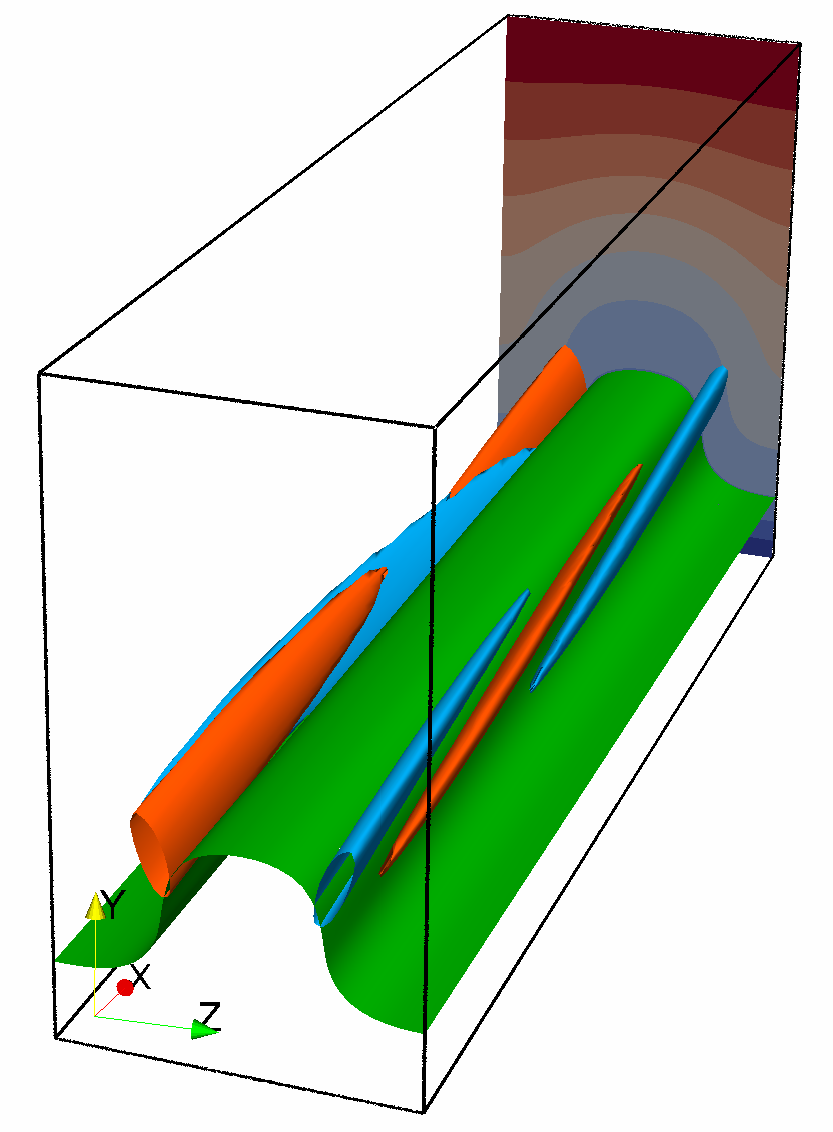}
    \caption{3D structure of the invariant near-wall solution at $Re=40,000$. The travelling wave is dominated by of a strong low-speed streak (situated in the centre of the figure) sandwiched between weaker high-speed streaks located closer to the wall and flanked by alternating mirror-symmetric pairs of counter-rotating vortices. 
     The low-speed streak is visualised by the green streamwise velocity isosurface ( $u=0.2 \, U_\infty$). Vortices are indicated by streamwise vorticity isosurfaces of $\omega_x$  at half of the maximum value with red/blue for positive/negative values. Due to the localisation of the structures near the wall, only the lower half of the computational domain is shown.}
    \label{fig:TW_ECS}
\end{figure}

\begin{figure}
    \centering
    {\includegraphics[width=0.6\textwidth]{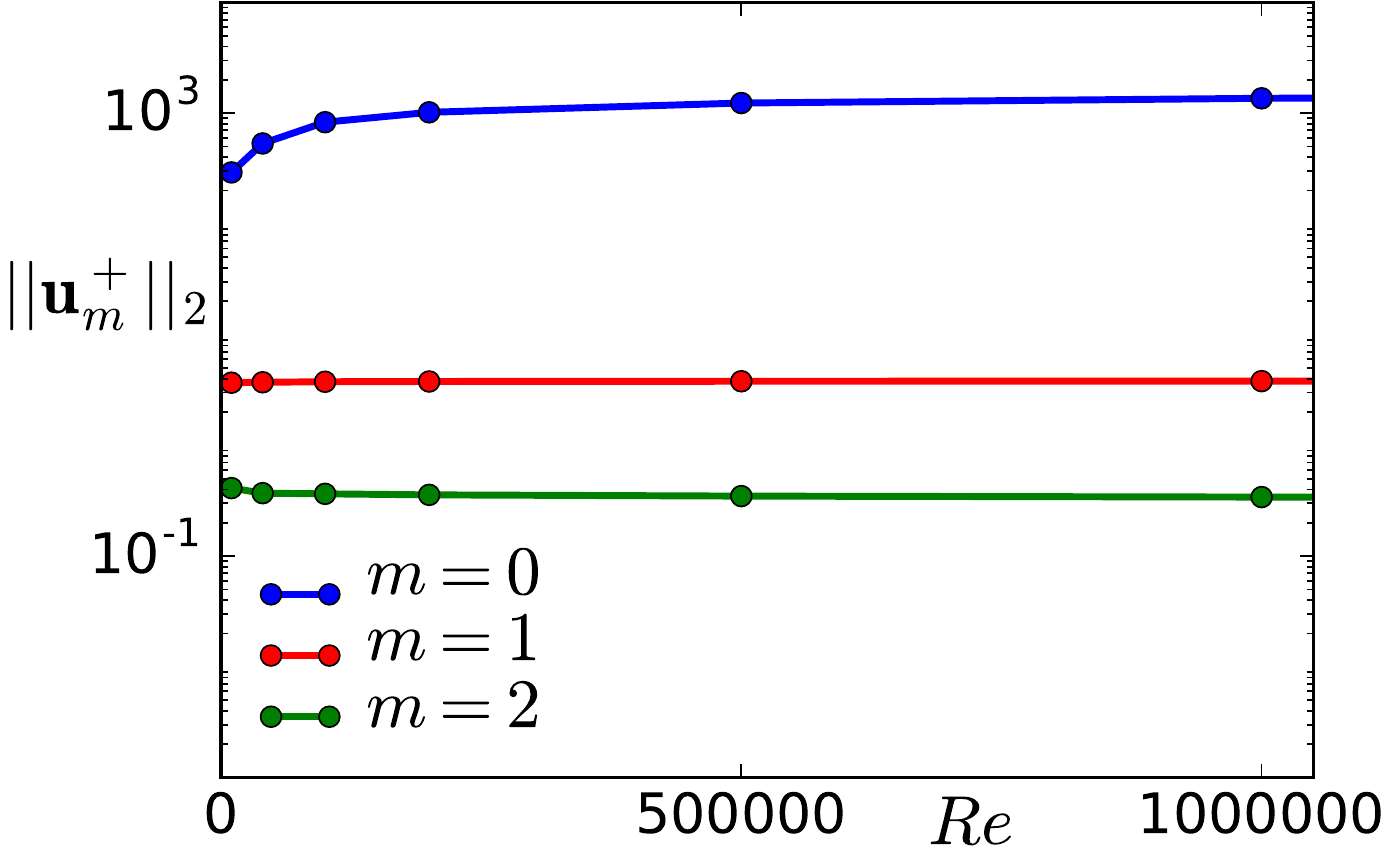}}
    \caption{Amplitude of the first three streamwise Fourier modes of the invariant solution expressed in inner units ($||\vel^+_0(y,z)||_2$, $||\vel^+_1(y,z)||_2$, and $||\vel^+_2(y,z)||_2$) as a function of $\Rey$.
    The amplitudes approach constant values indicating a solution that scales in inner units. The zero Fourier mode ($m=0$) dominates indicating a predominantly downstream invariant solution.}
    \label{fig:modes}
\end{figure}

%=================== FlowField at high Re
At high $\Rey$ the converged invariant solution remains localised near the wall. The structure of the travelling wave solution is dominated by spanwise-periodic alternating low and high-speed streaks flanked by counter-rotating vortical structures, as shown in \reffig{TW_ECS} where we present the solution at $\Rey=40,000$. The solution is dominated by streamwise oriented streaks. In \reffig{modes} we quantify the streamwise variation of the flow field by the amplitudes of the first three streamwise Fourier modes ($||\vel^+_0(y,z)||_2$, $||\vel^+_1(y,z)||_2$, and $||\vel^+_2(y,z)||_2$). For increasing $\Rey$ the inner-unit amplitudes approach constant values indicating a solution scaling in inner units. Moreover, the amplitude of the downstream independent zero mode is more than two orders of magnitude larger than the amplitude of the first mode and three orders of magnitude larger than the amplitude of the second mode. This confirms that the invariant solution is dominated by streamwise invariant structures.

\begin{figure}
    \centering
    \includegraphics[width=0.24\textwidth]{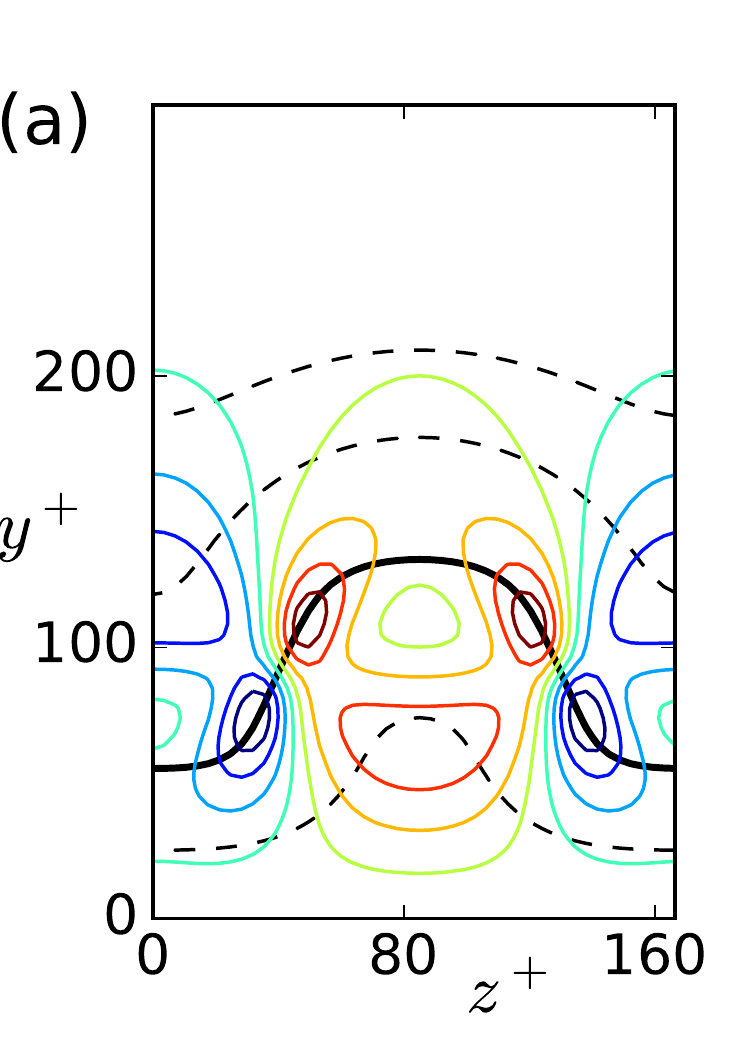}  
    \includegraphics[width=0.24\textwidth]{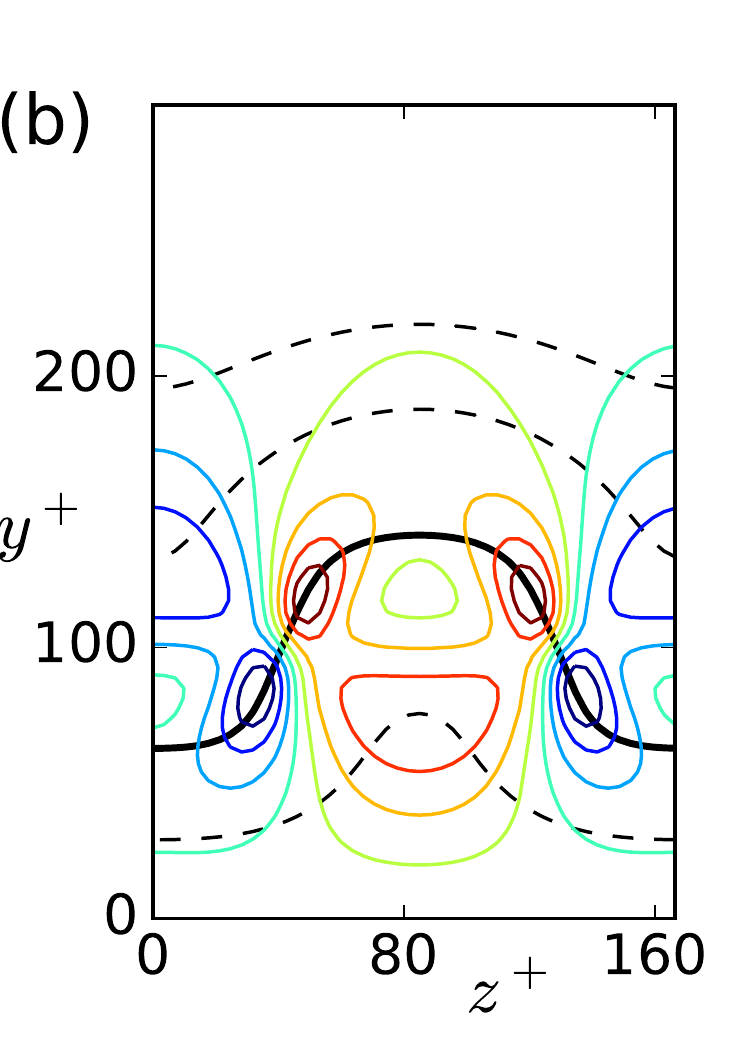}
    \includegraphics[width=0.24\textwidth]{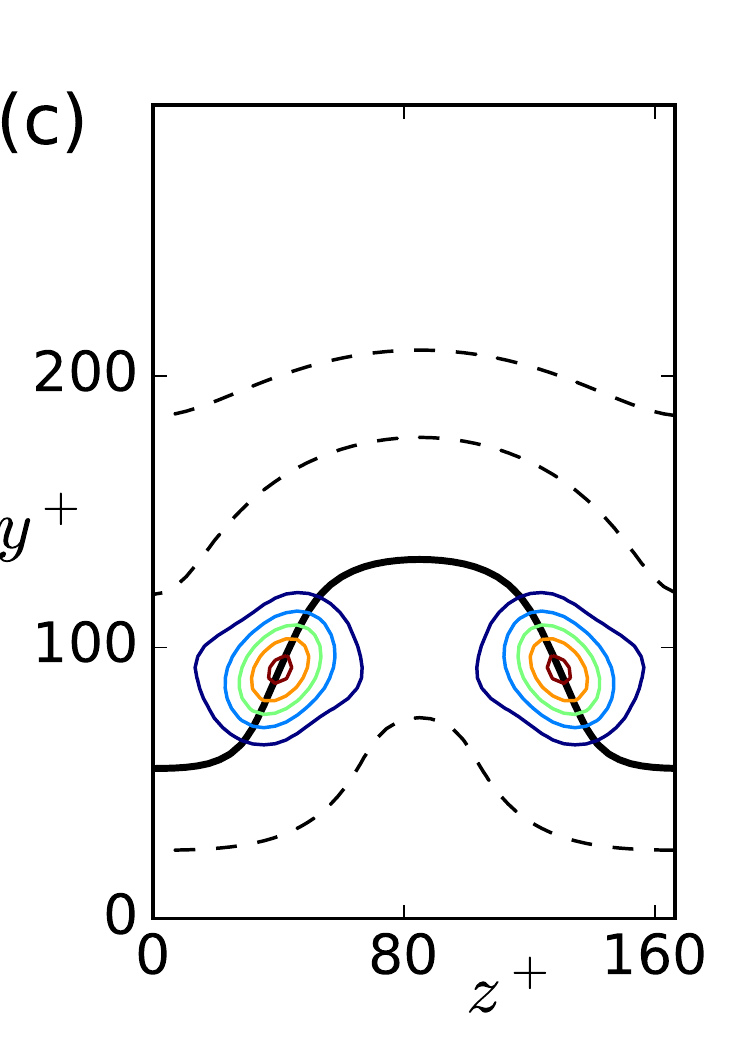} 
    \includegraphics[width=0.24\textwidth]{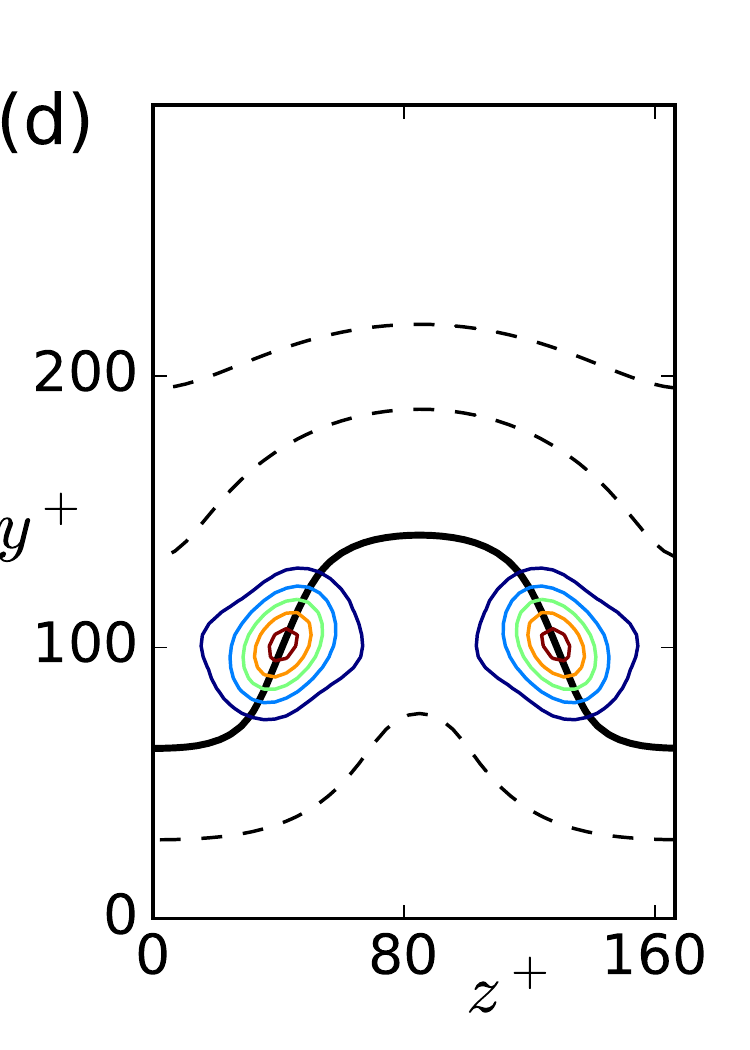}
    \caption{Contours of the streamwise Fourier modes of the invariant solutions' wall-normal velocity (coloured contour lines). The zero-th mode $v^+_0(y^+,z^+)$ (panels $a$ and $b$) and the first mode $v^+_1(y^+,z^+)$ (panels $c$ and $d$) shown in inner units at $\Rey=40,000$ (panels $a$ and $c$) and $\Rey=100,000$ (panels $b$ and $d$). 
The critical layer where the zero-th streamwise mode of the streamwise velocity equals the travelling wave's phase speed ($u_0(y,z)=c$) is also shown (bold solid black line) as well as the levels $u_0(y,z)/c=0.6, 1.4, 1.8$ (dashed black lines). Only the lower half of the numerical box is shown.
}
    \label{fig:CritLay}
\end{figure}

%=================== Characteristics
The relatively small bending of the low-speed streak (see \reffig{TW_ECS} and \reffig{modes}) of the identified near-wall invariant solutions are consistent with features of lower-branch solutions found at large scales in large domains with size typical of transitional \citep[e.g.][]{Wang2007} and turbulent large-scale motions \citep[e.g.][]{Rawat2016}.
In boxes remaining constant in outer units, this type of Navier-Stokes solutions assumes a critical-layer structure for high Reynolds numbers \citep{Wang2007, Deguchi2014,Deguchi2014b, Park2015} where the streaks' unstable mode concentrates near the critical layer. 
In the present case, however, the entire solution is downscaled in height, lateral wavelength and global amplitude when the Reynolds number increases (\reffig{cont_in}). There is no modification of the internal structure or a concentration near the critical layer. The zero-th streamwise Fourier mode of the wall-normal velocity $v^+_0(y,z)$ (associated to the streamwise vortices inducing the streaks), as well as the first streamwise Fourier mode $v^+_1(y,z)$ (associated to the streaks' instability mode) are asymptotically constant when expressed in inner units as $\Rey$ increases, as shown in \reffig{CritLay}.

\begin{figure}
    \centering
\includegraphics[width=0.33\textwidth]{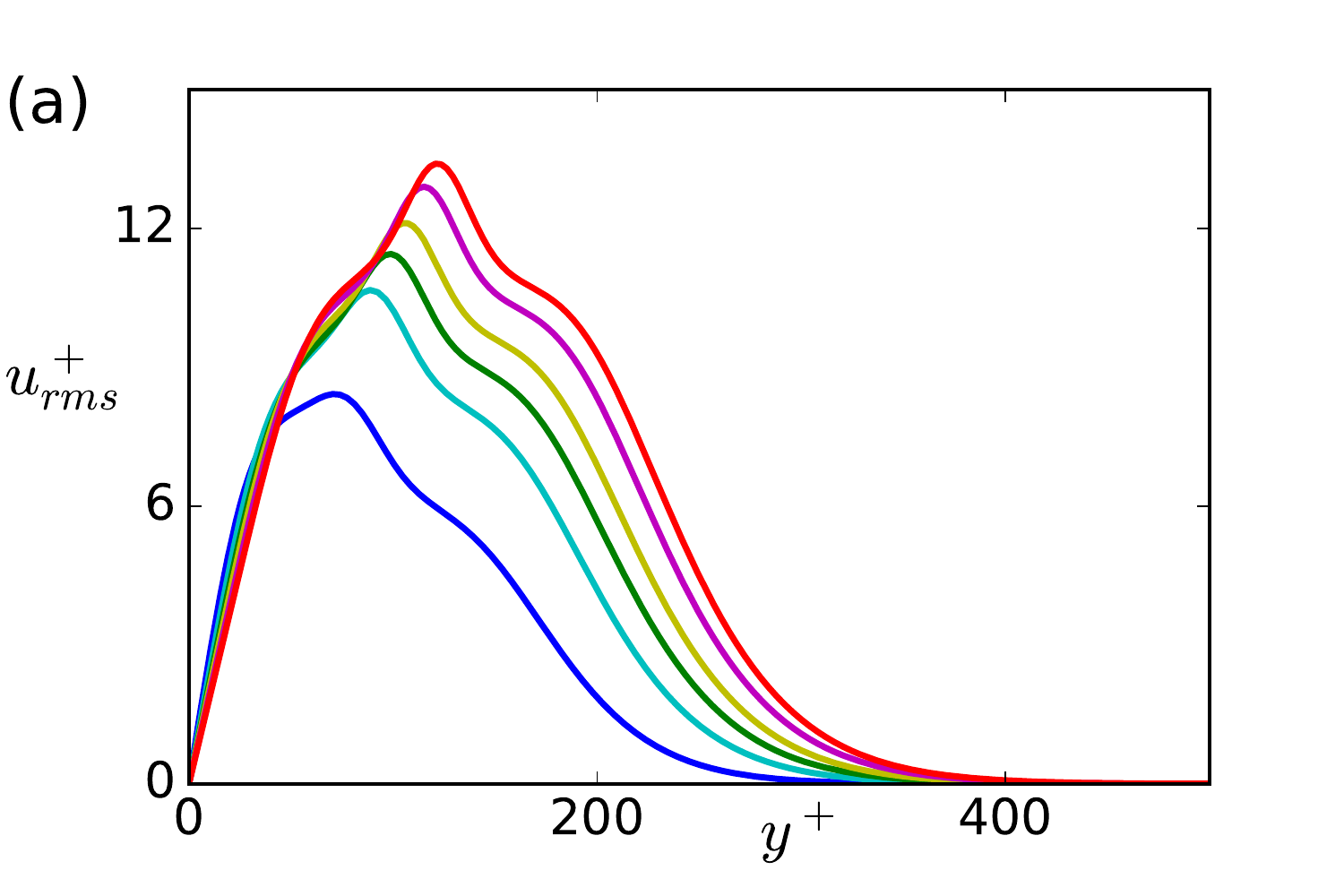} \hspace{-5mm}
\includegraphics[width=0.33\textwidth]{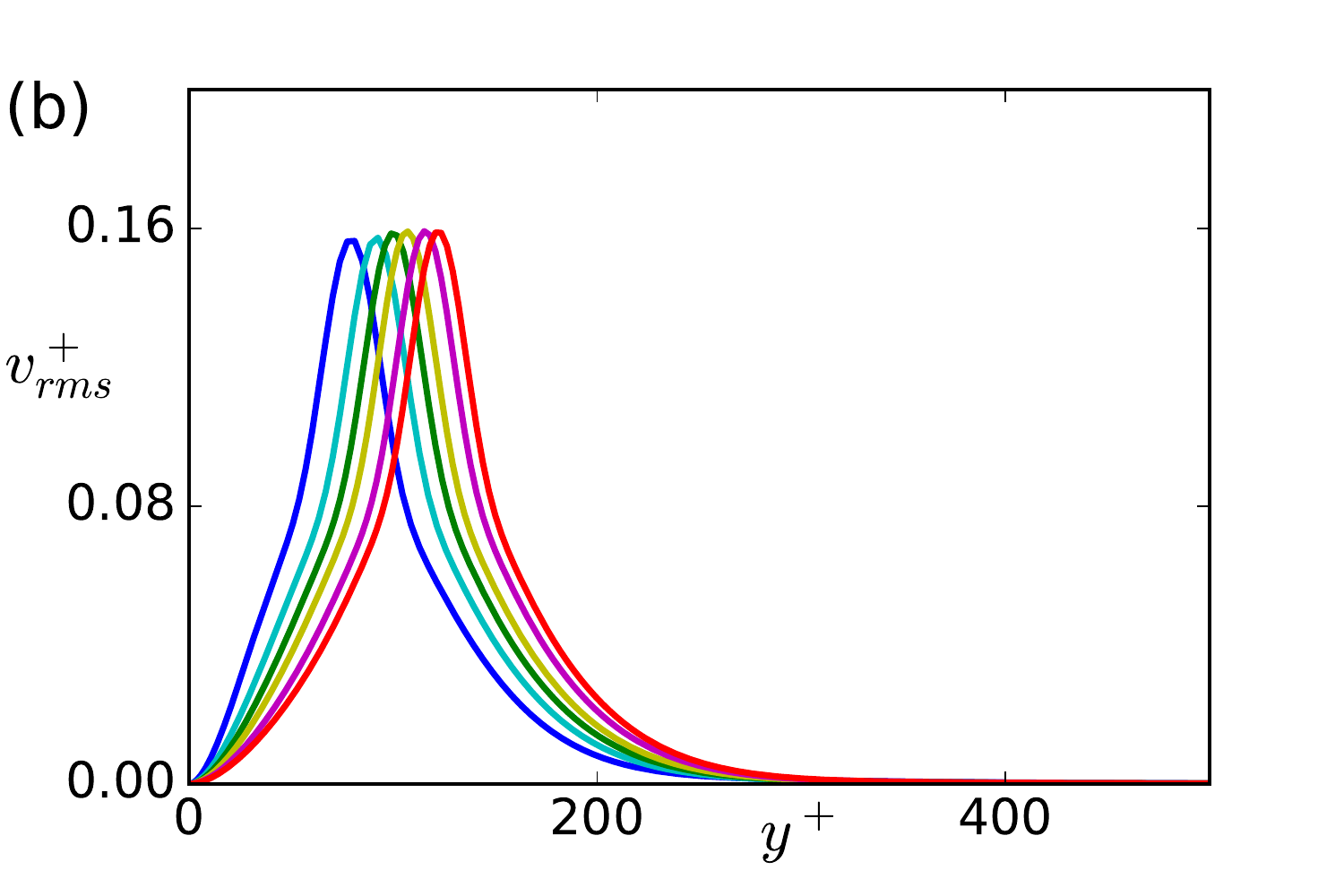} \hspace{-5mm} 
\includegraphics[width=0.33\textwidth]{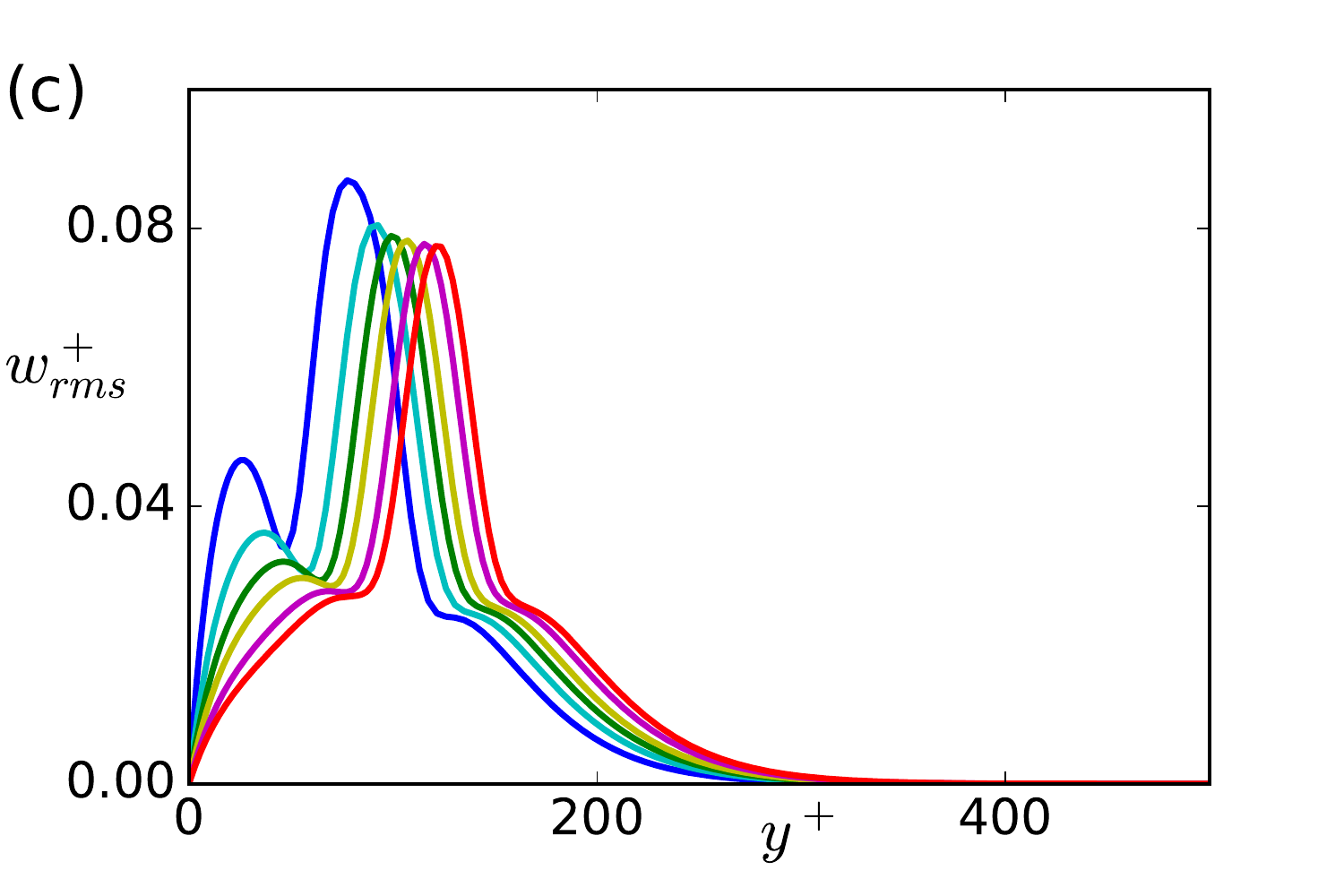} 
    \caption{Root mean squared (rms) profiles of the TW solutions expressed in inner units $u^+_{i,rms}(y^+)=\left[\frac{1}{L^+_x L^+_z}\int_0^{L^+_x}\int_0^{L^+_z} \left(u_i^+(x^+,y^+,z^+)-u^+_{i,mean}(y^+)\right)^2 dx^+ dz^+\right]^{1/2}$ at Reynolds numbers equal to $10,000$, $40,000$, $100,000$, $200,000$, $500,000$, and $1,000,000$. $(a)$ rms profiles of the streamwise velocity $u^+_{rms}$, $(b)$ rms profiles of the wall-normal velocity $v^+_{rms}$, $(c)$ rms profiles of the spanwise velocity $w^+_{rms}$. For increasing \Rey, the rms profiles asymptotically collapse indicating scaling in inner units. Note that the variation of $\Rey$ over two orders of magnitude corresponds to a full order of magnitude change in the height and lateral wavelengths of the solution in outer units (cf. \reffig{cont_in}).}
    \label{fig:TW_rms}
\end{figure}

% Add RMS
The root mean squared ($rms$) velocity profiles of the travelling wave solution expressed in inner units asymptotically collapse on a single curve when $\Rey$ is increased (\reffig{TW_rms}). This provides further confirmation that the travelling wave solution scales in inner units.

%=================== Asymptotic analysis
The small change of the solution with $\Rey$ (Fig. \ref{fig:CONTOUR} and \ref{fig:CritLay}), the asymptotically constant values of inner-unit amplitudes (Fig. \ref{fig:modes}), and the asymptotically converging rms profiles (\reffig{TW_rms})
provide strong evidence that the fully resolved travelling wave solution asymptotes towards a self similar solution at high $\Rey$.
To investigate if the solution indeed becomes independent of Reynolds number for $\Rey \to \infty$ when expressed in similarity variables defined by rescaled length and velocity scales, we 
use the friction velocity, $u_\tau=\frac{U_\infty}{\sqrt{Re}}$, and the viscous unit, $\delta_\tau=\frac{\delta^*}{\sqrt{Re}}$, to non-dimensionalise the evolution equations. The Navier-Stokes equation for the inner-unit velocity deviation from the laminar solution reads
\begin{align}\label{eq:NS_in}
\frac{\partial \vel^+}{\partial t^+}
+ \basevel^+ \cdot \nabla \vel^+
+ \vel^+ \cdot \nabla \basevel^+
+ \vel^+ \cdot \nabla \vel^+
= -\nabla p^+ + \nabla^2 \vel^+
\end{align}
where $\vel^+$ is the non-dimensionalised velocity deviation vector, and $\basevel^+$ is the non-dimensionalised laminar solution. The non-dimensionalised laminar solution in ASBL 
\begin{align}\label{eq:baseflow}
\basevel^+=\Rey^{1/2}\left(1-\exp \left(-y^+ \Rey^{-1/2} \right) \right) \mathbf{\hat{e}_x} - \Rey^{-1/2} \mathbf{\hat{e}_y}
\end{align}
is a function of the Reynolds number, $\Rey=\frac{U_\infty}{V_s}$, and the wall normal coordinate, $y^+$. The boundary conditions for the velocity deviation from the laminar base flow are periodic in the streamwise and in the spanwise directions and zero velocity at both the lower and upper wall. In the rescaled system with the given governing equation and the boundary conditions expressed in inner units, only the laminar base flow depends on $\Rey$. For $\Rey$ tending to infinity, the laminar base flow within the numerical box asymptotes to $\basevel^+=y^+ \mathbf{\hat{e}_x}$ and thus no longer depends on the Reynolds number. Therefore, when $\Rey$ is large, the rescaled system in inner units looses any dependence on the Reynolds number, and any solution of the system approaches a self-similar solution. Thus, any invariant solution that can be continued to asymptotically high Reynolds numbers in a box which has a fixed size in inner units becomes asymptotically self-similar.
This suggests that the invariant solution that we present in this paper represents a self-similar solution of the Navier-Stokes equations in the near-wall inner region of the asymptotic suction boundary layer flow at high Reynolds numbers.

The analysis shows that as $\Rey$ tends to infinity, the equations for ASBL solutions expressed in inner units loose the dependence on Reynolds number so that any solution of those equations is self-similar and scales in inner units. Remarkably, we also observe that the partial differential equations including boundary conditions that any wall-attached solution of ASBL satisfies at asymptotically high $\Rey$, are identical to those describing plane Couette flow (PCF) at a value of the typically used Couette Reynolds number $Re_{PCF}={H^+}^2/4$, based on half the gap height and half the velocity difference. Boundary conditions of ASBL enforce zero wall-parallel velocity and a non-zero wall-normal suction. Expressed in inner units the suction velocity is $V_s^+=\frac{1}{\sqrt{Re}}$. For large Re, suction effects thus vanish and asymptotically the standard no-slip boundary conditions of PCF are reached. Consequently, at asymptotically high $\Rey$, any wall-attached solution of ASBL corresponds to a wall-attached solution of PCF. The value of the Couette control parameter $Re_{PCF}$ formally depends on the arbitrarily chosen $H^+$ reflecting the fact that a solution localised at the wall, only depends on the shear rate at the wall while the distance of the second upper wall and thereby the value of $Re_{PCF}$ is irrelevant. Thus, at asymptotically high $\Rey$, all state space structures representing wall attached flow fields in the MFU of ASBL have counterparts in high Reynolds number PCF, such as those identified by \cite{Eckhardt2017}. This suggests that the relevant state-space structures for near-wall turbulence that we identified in ASBL are universal in that they are not only independent of $\Rey$ when expressed in inner units but also do not depend on the specific shear flow system considered. In fact, at sufficiently high $\Rey$, close to the wall, any wall-bounded shear flow is characterised by a universal shear profile, indistinguishable from PCF or ASBL and thus supports the same wall-attached solutions. This suggests that the entire state space of the near-wall region including the invariant solutions and their dynamical connections become independent of $\Rey$ and independent of the flow system. If invariant solutions, their heteroclinic connections and the entire state space structures are universal, the deterministic dynamics supported by those structures is also universal. This provides an explanation of the well-known fact that at sufficiently high flow speeds the statistics of near-wall turbulence becomes independent of the flow system.

%\clearpage
%%%%%%%%%%%%%%%%%%%%%%%%%%%%%%%%%%%%%%%%%%%%%%%%%%%%%%%%%%%%%%%%%%%%%%%%%%%%%%%%%%%%%%%%%%%%%%%
\section{Conclusion and discussion}\label{sec:concl}

The aim of this work is to demonstrate the existence of an exact invariant solutions of the Navier-Stokes equations that capture spatial scales typical of turbulent motions in the near-wall region of a boundary layer at high Reynolds numbers.
In a minimal flow unit of ASBL, chosen to capture the energetic scales of near-wall turbulent motions, 
a wall-attached travelling-wave solution has been computed by edge-tracking at $\Rey=1000$. We exploit the fact that ASBL allows to express the viscous length scale $\delta_\tau$ of the developed turbulent state in terms of the Reynolds number. We thus continued the solution to high $\Rey$ in the minimal flow unit, whose size remains constant in inner units but shrinks in outer units for increasing $\Rey$. The fully resolved solution can be followed up to $\Rey=1,000,000$. We provide numerical and theoretical evidence that the invariant solutions become exactly self-similar as $\Rey$ tends to infinity. Remarkably, the solution scales in inner units so that the individual fully resolved invariant solution of the Navier-Stokes equations captures the self-similar behaviour characteristic of near-wall-turbulent statistics. Moreover, in the high-$\Rey$ asymptotic limit, solutions of ASBL simultaneously constitute solutions of plane Couette flow thus reflecting the universality of near-wall turbulence.

Assuming that all relevant invariant solutions capturing near-wall motions can be continued to asymptotically high $\Rey$, our analysis suggests that entire state-space structures including invariant solutions and their dynamical connections become independent of Reynolds number when expressed in inner units. To provide further support for this picture, future research should aim at computing increasingly more complex state-space structures underlying near-wall turbulence in the high-$\Rey$ limit captured by the evolution equations expressed in inner units. This includes periodic orbits as well as orbits connecting invariant solutions.

Since the governing equations rescaled in inner units become asymptotically independent of Reynolds number, the complexity of the state space of near-wall turbulence and the number of relevant invariant solutions may not increase with $\Rey$ but remain constant, leading to a saturation of complexity in the near-wall region of turbulent flows. Such a saturation of complexity cannot be expected in the outer-region of turbulent flows. It may thus be possible to eventually provide a predictive and quantitative description of turbulence in terms of a manageable number of invariant solutions \citep{Chandler2013, chaosbook} not only for transitional flows but also for the universal near-wall region of wall-bounded turbulence at very high Reynolds numbers. The self-similar exact invariant solution in the near-wall region of a boundary layer reported here is a significant step towards extending the invariant solution approach to turbulence from transitional flows to near-wall region of fully developed boundary layers.

%%%%%%%%%%%%%%%%%%%%%%%%%%%%%%%%%%%%%%%%%%%%%%%%%%%%%%%%%%%%%%%%%%%%%%%%%%%%%%%%%%%%%%%%%%%%%%%
\appendix
\section{Methods}\label{app:methods}
We use the \texttt{ChannelFlow 2.0} code \citep{Gibson2019} to solve the nonlinear Navier-Stokes equations expressed in perturbation form
\begin{eqnarray}
\label{eq:NS}
\frac{\partial \vel}{\partial t}
+ \basevel \cdot \nabla \vel
+ \vel \cdot \nabla \basevel
+ \vel \cdot \nabla \vel
= -\nabla p + \frac{1}{\Rey} \nabla^2 \vel,
\end{eqnarray}
where $\vel$ is the perturbation relative to the laminar base flow $\basevel$ given by $\basevel=[U=1-e^{-y},V=-1/\Rey,W=0]$.
These momentum equations are complemented by the continuity equation $\nabla \cdot \vel=0$, periodic boundary conditions in the streamwise and in spanwise directions and the no-slip boundary conditions $\vel=\mathbf{0}$ at $y=0$ and $y=H$. To close the system, zero average pressure gradient in both streamwise and spanwise direction is imposed. The system of equations is discretised using a spectral collocation method based on Fourier-Chebychev-Fourier expansions in the streamwise, wall-normal and spanwise directions, respectively. The third-order accurate semi implicit backward differentiation method is used for time marching. Time steps are chosen such that the CFL number remains in the range of $0.4-0.6$.

For the large direct numerical simulations at $\Rey=333$, the large domain of size $L_x=243, H=225, L_z=121.5$ has been discretised with $N_x=256, N_y=301, N_z=256$ collocation nodes in the streamwise, wall-normal and spanwise directions, respectively. The simulation is initialised with a random field and after statistical steady state has been reached, the statistics are calculated from a time-series of $300,000$ advective time units.

The exact invariant travelling wave solutions have been computed using the Newton-Krylov-Hookstep method. We use a numerical resolution of $N_x=48$, $N_y=181$, and $N_z=48$ collocation points for the MFU with $L_x^+=633$, $L_z^+=170$, and $H^+=632$. Convergence is obtained when $||\left(\textbf{u}_{x-cT,y,z}|_{t=T}-\textbf{u}_{x,y,z}\right)||_{2}/T$ is less than $10^{-13}$, where $T=20$ and $||.||_2$ is not the energy norm of the velocity field but the L2-norm of the vector of independent Fourier-Chebychev coefficients of the operand. For $\Rey>100,000$ we carry out computations for the system expressed in inner units (\refeq{NS_in}). The solution vector itself has a typical magnitude $||\textbf{u}_{x,y,z}||_2$ on the order of $10$ so that the residual is approximately $14$ orders of magnitude smaller.

\begin{figure}
    \centering
    {\includegraphics[width=0.4\textwidth]{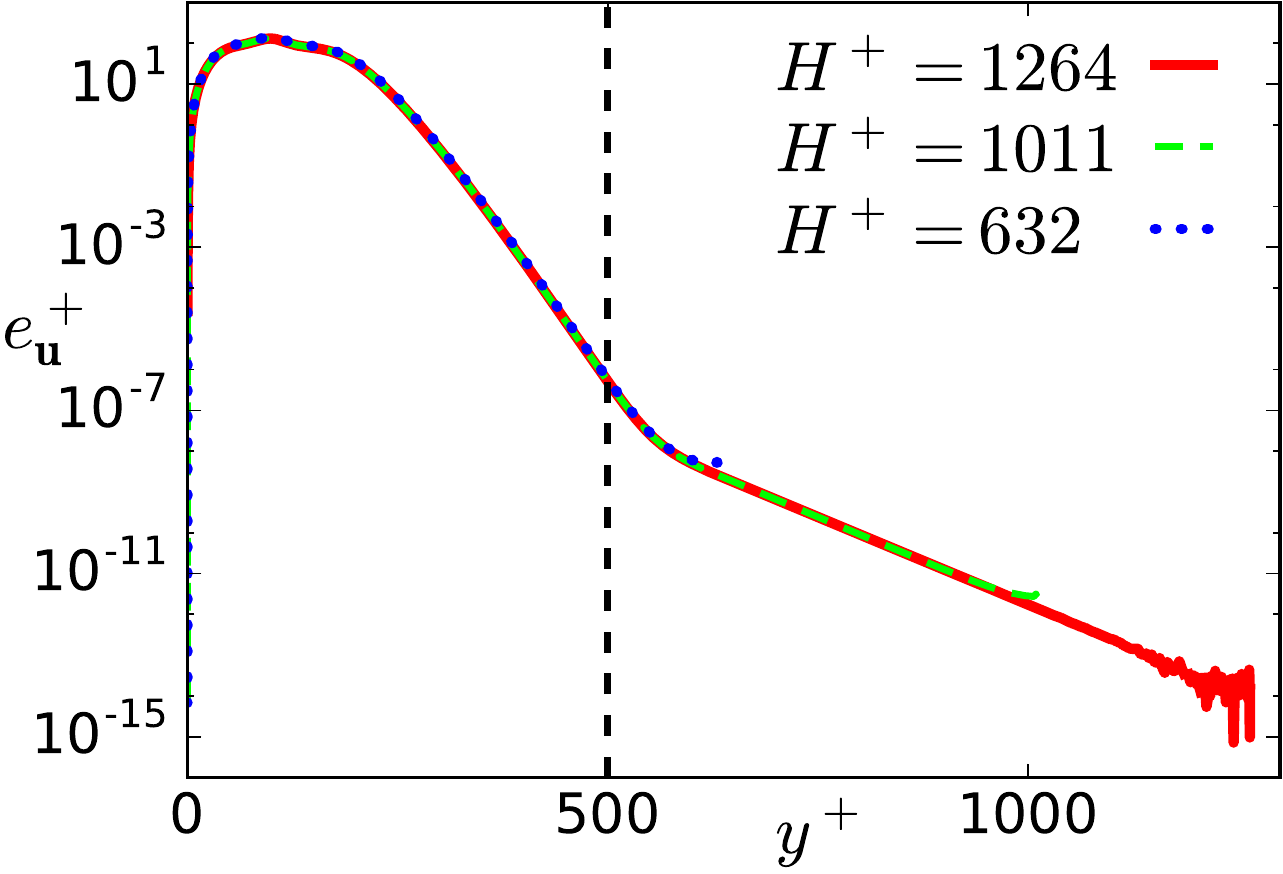}}
    \caption{Fluctuation energy profile of the solution expressed in inner units $e_\textbf{u}^+=u_{rms}^{+2}+v_{rms}^{+2}+w_{rms}^{+2}$ at $\Rey=100,000$ as a function of $y^+$ for three different box heights. The vertical dashed line show the height up to which rms plots are shown in \reffig{TW_rms}. At $y^+=500$, $e^+_\textbf{u}$ has dropped by nine orders of magnitude relative to its maximum.}
    \label{fig:H_indpn}
\end{figure}

The height of the MFU is chosen such that the solution is independent of $H^+$. This is confirmed in \reffig{H_indpn} where fluctuation energy profiles for three different box heights are depicted. As shown, increasing the box height from $H^+=632$ by $60\%$ and $100\%$ does not change the solution near the wall. As shown in \reffig{TW_rms}, the solution is well localised below $y^+=500$, the range over which rms profiles are plotted. At $y^+=500$, the energy has already dropped by nine orders of magnitude below its maximum. This confirms that our choice of $H^+=632$ is sufficiently large to ensure the solution is independent of $H^+$.

\begin{acknowledgements}
We thank Carlo Cossu for extensive discussions and comments on the manuscript. This work was supported by the Swiss National Science Foundation (SNSF) under grant no. 200021-160088. SA acknowledges support by the State Secretariat for Education, Research and Innovation SERI via the Swiss Government Excellence Scholarship.
\end{acknowledgements}

\bibliographystyle{jfm_abbrv}
\bibliography{references}
%\bibliography{./library}

\end{document}